\documentclass[%
 preprint,
 amsmath,amssymb,
 aps,
 prl,
]{revtex4-2}

\usepackage{graphicx}% Include figure files
\usepackage{dcolumn}% Align table columns on decimal point
\usepackage{bm}% bold math
\usepackage{graphicx}
\usepackage{siunitx}
\begin{document}

\preprint{APS/PRL}

%%TC:ignore

\title{Mesoscale Organization and Dynamics in Binary Ionic Liquid Mixtures}% Force line breaks with \\
%\title{Tuning Physicochemical Properties of Ionic Liquids: \\The Interplay of Mesoscale Aggregate Morphology and Dynamics}% Force line breaks with \\

%\title{Controlling transport and dynamics in ionic liquids using mesoscale organization  : \\The Interplay of Mesoscale Aggregate Morphology and Dynamics}% Force line breaks with \\

\author{Tyler Cosby}
\affiliation{Department of Chemical and Biomolecular Engineering, University of Tennessee, Knoxville, TN}
\author{Utkarsh Kapoor}
\affiliation{School of Chemical Engineering, Oklahoma State University, Stillwater, OK}
\author{Jindal K. Shah}
\email{jindal.shah@okstate.edu}
\affiliation{School of Chemical Engineering, Oklahoma State University, Stillwater, OK}
\author{Joshua Sangoro}
\email{jsangoro@utk.edu}
%\homepage{softmaterials.utk.edu}
\affiliation{Department of Chemical and Biomolecular Engineering, University of Tennessee, Knoxville, TN}

\date{\today}% It is always \today, today,

\begin{abstract}
The impact of mesoscale organization on dynamics and ion transport in binary ionic liquid mixtures is investigated by broadband dielectric spectroscopy, dynamic-mechanical spectroscopy, x-ray scattering, and molecular dynamics simulations. The mixtures are found to form distinct liquids with macroscopic properties that significantly deviate from weighted contributions of the neat components. For instance, it is shown that the mesoscale morphologies in ionic liquids can be tuned by mixing to enhance the static dielectric permittivity of the resulting liquid by as high as 100$\%$ relative to the neat ionic liquid components. This enhancement is attributed to the intricate role of interfacial dynamics associated with the changes in the mesoscopic aggregate morphologies in these systems. These results demonstrate the potential to design the physicochemical properties of ionic liquids through control of solvophobic aggregation.% achieved by  mixing neat ionic liquids.
% The mixtures approach opens up new parameter space for "novel" ILs with properties that are inaccessible to the neat systems.
\end{abstract}

% The impact of mesoscale organization and dynamics on macroscopic ion transport in ionic liquids is investigated 
%%TC:endignore

\maketitle

%\tableofcontents

Elucidating the influence of mesoscale organization on the dynamics and transport properties of ionic liquids is critical to developing design criteria for their applications in chemical synthesis, nanoparticle growth, biomass processing, batteries, solar cells, and supercapacitors.\cite{Hallett:2011aa,He:2015aa,Antonietti,MacFarlane:2014aa,Chakrabarti:2014aa,Smiglak:2014aa,Armand2009,Osch:2017aa,Santos:2011aa,Santos:2011ab} In the past decade, the formation of mesoscale polar and non-polar domains in ionic liquids with substantial non-polar, alkyl side groups was recognized in detailed x-ray scattering, neutron scattering, and molecular dynamics (MD) simulation studies.\cite{Hayes:2015aa,Russina:2017aa,Canongia-Lopes:2006aa,Triolo:2007aa, Kapoor:2017, Kapoor:2018} Mesoscale organization has been used to qualitatively explain numerous experimental findings which imply spatially and temporally distinct regions within bulk ionic liquids.\cite{Araque2015,Hayes:2015aa,Berrod:2017aa,Ferdeghini:2017aa,Burankova:2017aa,Aoun2010,Griffin:2016aa,Wang2006,Sonnleitner:2014aa,Turton:2009aa,Fayer:2014aa} Recent studies suggest that the existence and dynamics of the aggregates in neat ionic liquids, associated with fluctuations of the polar and non-polar regions, correlate strongly with many of the physicochemical properties of ionic liquids including transport properties such as zero-shear viscosity, dc ionic conductivity, and static dielectric permittivity.\cite{Cosby2017a,Yamaguchi2018,Russina:2017ab,Russina:2009aa,Kofu:2013aa,Castner:2010aa} However, no efforts to exploit the mesoscale organization to design novel ionic liquids with unique physical and chemical properties have been reported. In this work, it is demonstrated that by mixing ionic liquids with varying degrees of solvophobic aggregation, it is feasible to design distinct liquids with macroscopic properties that significantly deviate from weighted contributions of the neat components.

The local organization, or morphology, of the mesoscale aggregates is in part determined by the relative volume fractions of the polar and non-polar groups of the component ions.\cite{Hayes:2015aa} In amphiphilic imidazolium, pyrrolidinium, piperidinium, quaternary phosphonium, and quaternary ammonium ionic liquids, increasing the alkyl chain length on the cation headgroup tends to swell the non-polar domain leading to a progression from globular morphology to a loosely-defined bicontinuous morphology with percolating polar and non-polar domains.\cite{Shimizu:2014aa,Canongia-Lopes:2006aa,Triolo:2007aa,Shimizu:2010aa} Conversely, at a given length of the alkyl chain on the cation, the polar domain may be enlarged (or reduced) by selecting an anion with a larger (or smaller) molar volume.\cite{Hayes:2015aa} An alternative approach to altering the polar and non-polar volume fractions is to consider mixtures of two or more ionic liquids with differing chemical structures.\cite{Russina:2017ab} We hypothesize, for instance, that in an amphiphilic ionic liquid mixed with a predominantly polar ionic liquid, which preferentially locates within the polar domain, the polar \textit{versus} non-polar volume fraction and accordingly the morphology can be tuned by simply altering the composition. Any number of ion chemical structure combinations can be envisioned for the mixture components. Additional and more complex morphologies can be accessed by mixing ionic liquids with differing chain lengths as well as other chemical structure features.\cite{Russina:2017ab} 
Totally unexplored is the influence of composition-dependent morphology and the accompanying mesoscale aggregate dynamics on the physical and chemical properties of ionic liquid mixtures. Unraveling this interplay requires the systematic investigation of well-chosen mixtures by a combination of techniques capable of probing mesoscale aggregate morphology, ion and mesoscale aggregate dynamics, as well as the bulk or macroscopic transport properties. 
 
 In this study, we highlight an approach in which complementary experimental and computational techniques are employed to investigate changes to mesoscale aggregate morphology and dynamics as a function of composition in binary mixtures of the ionic liquids 1-octyl-3-methylimidazolium tetrafluoroborate (C$_8$MIm BF$_4$) and 1-ethyl-3-methylimidazolium tetrafluoroborate (C$_2$MIm BF$_4$). We find that by mixing these two imidazolium ILs, which differ only in the cationic alkyl chain length, we can transform the bicontinous morphology of neat C$_8$MIm BF$_4$ to more isolated and spherical non-polar aggregates as indicated by x-ray scattering and molecular dynamics simulations. As a result of the composition-dependent evolution in morphology, the mesoscale aggregate dynamics, as probed by dynamic-mechanical and broadband dielectric spectroscopy, are significantly altered. The changes to aggregate morphology and dynamics result in a 100$\%$ increase in the static dielectric permittivity, also known as the dielectric constant, relative to that of either pure component. 

  The chemical structures of C$_8$MIm BF$_4$ and C$_2$MIm BF$_4$ were chosen to approximate the desired mixture consisting of one amphiphilic and one predominantly polar ionic liquid. The imidazolium head group and the anion of each ionic liquid is identical in order to minimize the potential influence of mixture composition on the ion-ion interactions within the polar domains and at polar/non-polar interfaces. In this way, the influence of mesoscale aggregate morphology and dynamics on the transport properties may be investigated independent of any change in ion-ion interactions. To probe the influence of composition on the mesoscale aggregate morphology, and to verify the location of C$_2$MIm BF$_4$ within the polar domain, neat C$_8$MIm BF$_4$, neat C$_2$MIm BF$_4$, and 30, 50, and 70mol$\%$ C$_2$MIm BF$_4$ mixtures were investigated by x-ray scattering and molecular dynamics simulations. 
  
\begin{figure}
    \centering
   \includegraphics{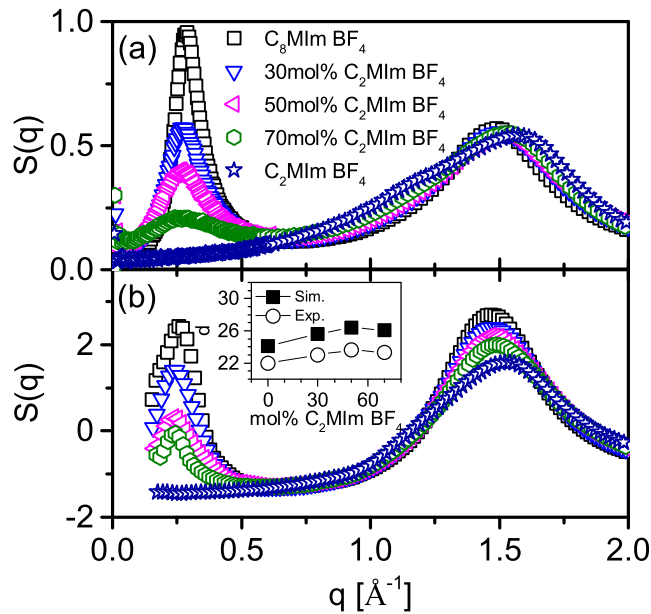}
    \caption{(a) Structure factors, S(q), obtained by x-ray scattering. (b) Structure factors computed by MD trajectories. Inset: Comparison of the real-space correlation distance, $d=2\pi/q_{max}$, of the pre-peak obtained by simulation (sim.) and experiment (exp.).
}
    \label{fig: Figure1}
\end{figure}

 The structure factors, S(q), of the neat ionic liquids C$_{8}$MIm BF$_{4}$ and C$_{2}$MIm BF$_{4}$ and the 30, 50, and 70mol$\%$ C$_{2}$MIm BF$_{4}$ mixtures obtained at room temperature by small- and wide-angle x-ray scattering are presented in Figure \ref{fig: Figure1}(a). C$_{8}$MIm BF$_{4}$ exhibits a pre-peak at q = \SI{0.28}{\per\angstrom} typical of self-assembled ionic liquids and assigned to the scattering from polar domains separated by a non-polar domain.\cite{Hayes:2015aa} The higher q-peak arises from adjacency correlations of both inter- and intramolecular origins and is common to all ionic liquids.\cite{Araque2015} C$_{2}$MIm BF$_{4}$ has no pre-peak and is therefore taken to be non-aggregating. With increasing concentration of C$_{2}$MIm BF$_{4}$ the pre-peak is reduced in intensity and shifts to slightly lower q-values. Insight into the structural changes which alter the position and intensity of the pre-peak is provided by complementary MD simulations. Details of the structure factors calculated from the MD simulations may be found in the Supplementary Information. The structure factors, shown in Figure \ref{fig: Figure1}(b), reproduce the positions and relative intensities of the experimental structure factors reasonably-well over the entire q-range. The real space distances corresponding to the pre-peak, $d=2\pi/q_{max}$, found by experiments and simulations are presented in the inset of Figure \ref{fig: Figure1}(b). MD simulations slightly over-predict the experimental values, however, the non-monotonic dependence of the domain distance on composition is well-reproduced. The excellent agreement between MD simulation and experimental results provides confidence in the assignment of certain composition-dependent morphological transitions which are described by the subsequent detailed analysis of the MD simulations.

\begin{figure}
     \centering
     \includegraphics{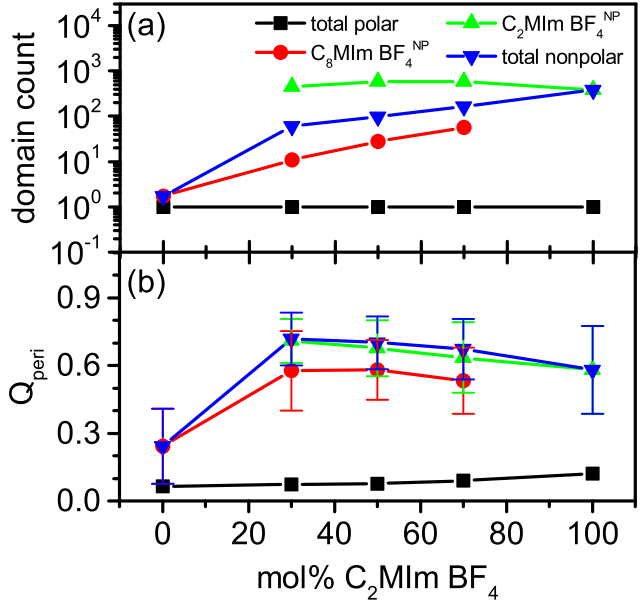}
     \caption{(a) Average domain count of the polar and non-polar domains present in the simulation box as a function of C$_{2}$MIm BF$_{4}$ concentration. (b) Average isoperimetric quotient, Q$_{\textrm{peri}}$, of respective cation/anion polar and nonpolar domains as a function of C$_{2}$MIm BF$_{4}$ concentration. Please note that average numerical value and standard deviations were computed by dividing the trajectory into three blocks. }
     \label{fig:Figure2}
 \end{figure}

The connectivity of the nano-segregated polar/non-polar structure is examined in a quantitative manner in terms of domain analysis based on Voronoi tessellation technique.\cite{Brehm2011} In this analysis, adjacent Voronoi cells sharing a face and belonging to a given subunit constitute a domain. For our purposes, each of the binary ionic liquid mixture systems is characterized in terms of four unique domains: (a) the total polar domain composed of the polar groups of both the cations and the anion; (b) C$_8$MIm non-polar; (c) C$_2$MIm non-polar and (d) total non-polar containing the non-polar groups from both cations. The polar group of both imidazolium cations contains the imidazolium ring as well as the methyl and methylene groups directly bonded to the ring, while the anion is completely polar. The polar group of the cation and anion together constitutes the overall polar domain. The non-polar regions in the two cations are the respective uncharged carbon groups minus the methylene group directly bonded to the imidazolium ring. The uncharged alkyl chain of the cations are considered unique in order to identify the origin of the structural changes at various concentrations. Figure \ref{fig:Figure2}(a) provides number of domains based on this classification as a function of the C$_2$MIm BF$_4$ concentration.  As expected, a domain count of 1 is observed for the polar domain indicating its three-dimensional connectivity for all the ionic liquids mixtures studied here. This observation is in line with previous simulation studies involving a wide range of pure ionic liquids.\cite{Shimizu:2014aa,Bernardes2014, Shimizu2015,Shimizu2017a,Brehm2015, Brehm2011, Hayes:2015aa, Kapoor:2017, Kapoor:2018} For pure C$_2$MIm BF$_4$, the domain counts for the non-polar group are significantly higher than 1 ($\sim$ 380) indicating that the cation non-polar groups are dispersed in the system. On the other hand, the domain count for the non-polar tails in the pure C$_8$MIm BF$_4$ ionic liquid is between 1 and 2, indicating that the majority of alkyl chains are connected in a single percolated non-polar domain with some possible occurrence of isolated C$_8$MIm BF$_4$ non-polar chains. The addition of 30mol$\%$ C$_2$MIm BF$_4$ results in a significant disruption of the non-polar connectivity as the large single continuous domain is broken into as many as 10 separate domains and the number reaches as high as 57 at the highest C$_2$MIm BF$_4$ concentration.

A quantitative metric of the variety of shapes adopted by the polar and non-polar domains is provided by the isoperimetric quotient, $Q_{\textrm{peri}}=\left[r_{\text {sphere}}(V)/r_{\text {sphere}}(A)\right]^{6}=36\pi\left[V^{2}/A^{3}\right]$, where $V$ and $A$ denote the volume and area of a given domain respectively while $r_{\text{sphere}}(V)$ and $r_{\text{sphere}}(A)$ represent the equivalent radii of the sphere with volume $V$ and the sphere with area $A$, respectively.  With this definition, the shape parameter will assume a value of 1 for a perfectly spherical shape while any deviations from sphericity lead to the values lower than 1.\cite{Brehm2015} The change in the isoperimetric quotient as a function of the C$_2$MIm BF$_4$ concentration is shown in Figure \ref{fig:Figure2}(b). From the figure, it is clear that Q$_{\textrm{peri}}$ for the polar domain shows a negligible dependence on the concentration of C$_2$MIm BF$_4$ and is always less than 0.1, which implies that the shape of the polar network differs greatly from sphericity. Further, the non-polar domain present in pure C$_8$MIm BF$_4$ ionic liquid has a Q$_{\textrm{peri}}$ value less than 0.25, and domain count of approximately 1, suggesting a network whose shape is also far from spherical. However, with the introduction of 30 mol$\%$ of C$_2$MIm BF$_4$ in C$_8$MIm BF$_4$, the Q$_{\textrm{peri}}$ value nearly doubles assuming a value of $\sim$ 0.58 suggesting a transition in the morphology of the domains which now more closely resemble a sphere in comparison to that in the pure C$_8$MIm BF$_4$. The results are even more dramatic when the total non-polar domain of the mixture is considered with values approaching as high as 0.7 at 30 mol\% C$_2$MIm BF$_4$. Above 30mol$\%$, the Q$_{\textrm{peri}}$ is practically composition independent, indicating that the transition in mesoscale aggregate shape occurs at or below this concentration.
We conjecture that the dispersed subphase of C$_2$MIm modulates the overall non-polar domain connectivity and morphology in a way that alters the dynamics of the mesoscale aggregates and gives rise to a variation in the static dielectric permittivity as a function of C$_2$MIm BF$_4$ concentration, as discussed later in the current work. 

 \begin{figure}
    \centering
    \includegraphics{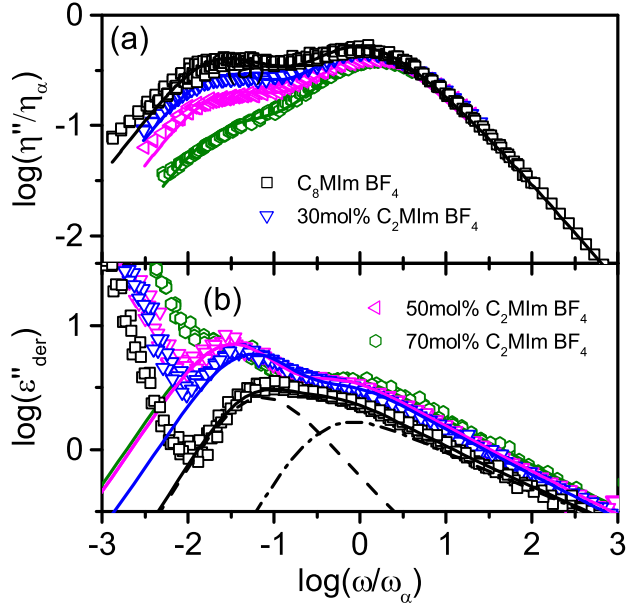}
    \caption{(a) Imaginary part of complex viscosity, $\eta^*=\eta'-i\eta''$, normalized by the structural relaxation contribution to the zero-shear viscosity, $\eta_{\alpha}=G_{\infty}/\omega_{\alpha}$, \textit{versus} frequency normalized by the structural relaxation rate, $\omega_{\alpha}$. Solid lines correspond to the total fit by two Cole-Davidson-modified Maxwell relaxation models. (b) The derivative representation of the dielectric loss, $\varepsilon''_{der}$. T=204-228K at 8K increments. Solid lines correspond to the total fit of two Havriliak-Negami fit functions at 204K. The dashed and dotted-dashed lines represent the separate Havriliak-Negami functions of the underlying slow and structural relaxations, respectively, for C$_8$MIm BF$_4$. Details of the fitting functions and their parameters are provided in the Supplementary Information.}
    \label{fig: Figure3}
\end{figure} 
 
 To probe the influence of composition-dependent morphology on the  mesoscale aggregate dynamics, the neat ionic liquids and their mixtures were investigated by dynamic-mechanical and broadband dielectric spectroscopy. These data are presented in Figure \ref{fig: Figure3} in terms of the imaginary part of complex viscosity, $\eta''$, and the derivative representation of the dielectric loss, $\varepsilon''_{der}$, respectively. Two distinct relaxation peaks, well separated in frequency, are observed in each experiment. The composition-dependent calorimetric glass transition temperatures, T$_g$'s, obtained by differential scanning calorimetry, follow a weighted average of the two neat IL components.  The rates of the faster, higher-frequency mechanical and dielectric relaxations scale by T$_g$. This comparison indicates that, in both experiments, the faster relaxations arise from the same underlying ion dynamics which define the glass transition in this class of ionic liquids.\cite{Sangoro:2012ab,Russina:2009aa} By comparison, the slower dielectric relaxation rate, $\omega_{slow, BDS}$, has a more complex composition dependence, as discussed later, indicating it is not directly associated with the glass transition as was recently suggested.\cite{Pabst2019} Further details on the T$_g$'s and temperature-dependent relaxation rates are presented in the Supplementary Information. In order to emphasize the influence of composition on the slower mesoscale dynamics, the spectra in Figure \ref{fig: Figure3} are shown \textit{versus} frequency normalized by the rate of the structural relaxation, $\omega_{\alpha}$. In this representation, it is evident that an additional relaxation is present in each experiment at rates slower than that of the primary structural relaxation in agreement with prior studies.\cite{Cosby2017a,Cosby:2018aa,Griffin:2015aa}
 
In neat C$_{n}$MIm BF$_{4}$ ionic liquids, the emergence of the slow dynamics was found to coincide with the onset of solvophobic aggregation, as evidenced by the development of the x-ray scattering pre-peak, and by a comparison between the relaxation rates with those previously obtained by neutron spin echo spectroscopy.\cite{Cosby2017a,Russina:2017ab,Russina:2009aa,Kofu:2013aa} Therefore, the slow relaxations were attributed to fluctuations of the mesoscale aggregates at timescales longer than the structural relaxation. This attribution is further substantiated by Yamaguchi's recent computational work which shows that a cross-correlation exists between the shear stress relaxation and the slow relaxation of the domain structure corresponding to the scattering pre-peak.\cite{Yamaguchi2018} It should also be noted that similar slow, sub-structural relaxations are also observed in some other mesoscopically ordered liquids, most notably monohydroxy alcohols, where they are also attributed to a supramolecular origin.\cite{Gainaru:2014ac,Hecksher:2014aa,Hecksher:2016aa,Arrese-Igor2018,Bohmer:2014ac,Yamaguchi2018a} Despite the apparent similarities in the dielectric and dynamic-mechanical spectra, these techniques are sensitive to distinctly different correlations within the bulk liquid, i.e. polarization and the stress tensor, respectively. Therefore, a careful comparison of the influence of composition on the strength and rate of the slower mesoscale aggregate dynamics, as obtained by each technique, may provide a useful insight into its molecular origin. 
 
Upon dilution of C$_8$MIm BF$_4$ with C$_2$MIm BF$_4$, the slow mechanical mesoscale relaxation is gradually reduced in strength until at 70mol$\%$ C$_2$MIm BF$_4$ it is barely visible as a low-frequency shoulder to the structural relaxation. However, relative to the structural relaxation rate, the rate of the slow relaxation is independent of composition. This trend is in stark contrast to observations of the slow dielectric relaxation. Upon addition of C$_2$MIm BF$_4$, the slow dielectric relaxation substantially increases in strength and shifts to lower frequencies relative to the structural, $\alpha$-relaxation rate. The relaxation rates and strengths are provided in the Supplementary Information. The divergence in the composition dependence of the slow relaxations probed by each technique indicates a possible sensitivity of the dielectric relaxation mechanism to the mesoscale aggregate shape or morphology which is lacking in the mechanical relaxation. 

Several factors could potentially influence the mesoscale aggregate dynamics such as the composition-dependent volume fraction, shape, and size of the aggregated non-polar domains as well as any alteration in ion-ion interactions at the polar/non-polar interfaces. By our choice of cation and anion for the two ionic liquid components we have attempted to minimize the latter effect and will not consider it further. The size of the non-polar domains is probed by the real-space correlation distance corresponding to the x-ray scattering pre-peak. These distances, given in the inset of Figure \ref{fig: Figure1} (b), increase with increasing concentration of C$_{2}$MIm BF$_{4}$. However, the modest increase in aggregate dimensions by $\approx$ \SI{2}{\angstrom} is not considered sufficient to explain the substantial changes in aggregate dynamics observed in the mechanical and dielectric spectra. We also note the relative invariance of the mechanical mesoscale aggregate relaxation rate as evidence for this relatively minor change in non-polar domain size. The gradual reduction in the strength of the slow mechanical relaxation is consistent with a reduction in volume fraction of the non-polar aggregate domains in which it originates. Accordingly, the opposite composition-dependence of the slow dielectric relaxation strength suggests an overriding influence of aggregate shape rather than volume fraction.

 \begin{figure}
    \centering
    \includegraphics{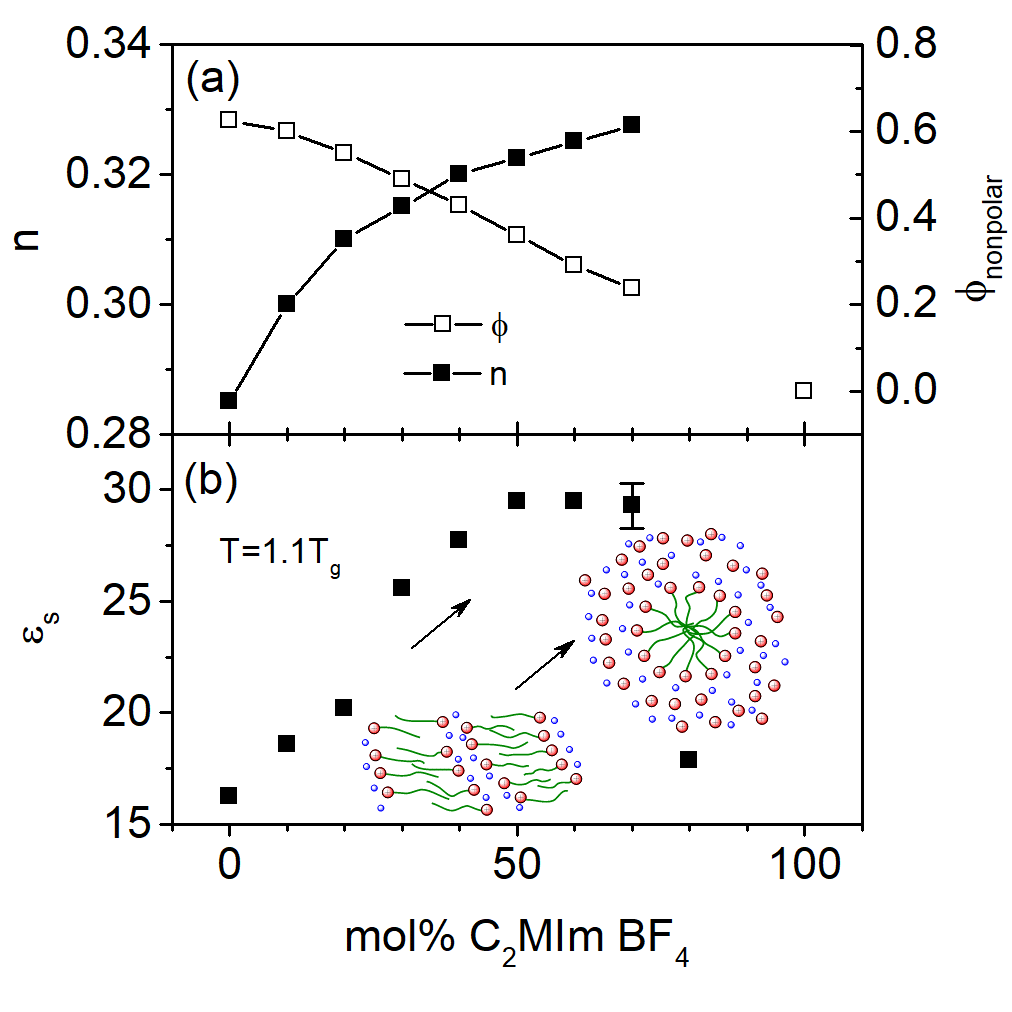}
    \caption{(a) Shape parameter, n, and volume fraction of  the non-polar phase, $\phi_{\textrm{nonpolar}}$, \textit{versus} mol$\%$ C$_{2}$MIm BF$_{4}$. (b) Concentration dependence of static dielectric permittivity at T$=1.1$T$_{g}$. The increase in $\varepsilon_{s}$ is due to the concentration-dependent aggregate shapes illustrated by the inset cartoons.}
    \label{fig: Figure4}
\end{figure} 

It is proposed that the slow dielectric relaxation originates from interfacial polarization at the polar/non-polar interfaces. From numerous studies on heterogeneous liquids and solids it is well established that interfacial polarizations are strongly dependent on the shapes of the included domains.\cite{Torquato:2002aa,Choy:1999aa,Looyenga:1965aa,Beek:1967aa} A change in the shapes of the aggregates might therefore be the origin of the observed increases in strength of the slow relaxation. The influence of a transition in aggregate shape on interfacial polarization can be ascertained using an effective medium approximation (EMA). EMAs are useful approximate approaches to relate the shape and volume fractions of a filler phase located within host matrices to the macroscopic dielectric properties of the composite, provided the properties of the two phases can be estimated.\cite{Boyle:1985aa,Kremer:2003aa,Beek:1967aa,Mizoshiri2010} Insight into the interplay of aggregate shapes, volume fractions, and dielectric relaxation strengths may be obtained by probing the ability of an EMA to accurately predict the static dielectric permittivities, $\varepsilon_{s}$, and dc ionic conductivities, $\sigma_{0}$, of the mixtures. These values are defined as the low-frequency limiting values of the real parts of complex dielectric permittivity and complex conductivity, respectively, see Supplementary Information.\cite{Kremer:2003aa} The static dielectric permittivity, $\varepsilon_{s}=\Delta\varepsilon_{slow}+\Delta\varepsilon_{\alpha}+\varepsilon_{\infty}$, contains contributions from all higher frequency dielectric relaxations including the structural and slow dielectric relaxation strengths, $\Delta\varepsilon_{\alpha}$  and $\Delta\varepsilon_{slow}$, as well as other processes included in the high-frequency limiting permittivity, $\varepsilon_{\infty}$. In the ionic liquid mixtures, the composition dependence of $\varepsilon_{s}$ is dominated by and is roughly proportional to the strength of the slow dielectric relaxation. For our investigation, we employ a form of the symmetric Looyenga equation\cite{Boyle:1985aa}, which is suitable for the conducting phases and intermediate volume fractions found in our IL mixtures.\cite{Boyle:1985aa} Details of this EMA and our application of it may be found in the Supplementary Information. The two fit parameters of this model, $n$ and $\phi$, correspond to the shape factor and volume fraction of the non-polar domain, see Figure \ref{fig: Figure4}(a). The experimental values of $\varepsilon_{s}$ and $\sigma_{0}$ are predicted only by an increase in $n$ and a concomitant decrease in $\phi$. Due to the assumptions on which an evaluation of the EMA relies, these trends can only be interpreted qualitatively. The overall reduction in volume fraction of the non-polar domain is consistent with its dilution upon addition of the non-aggregating C$_{2}$MIm BF$_{4}$. The shape factor, $n$, is related to the shape of the insulating phase, with $n$ $<$ $1/3$ corresponding to rod-like inclusions and n $=1/3$ to spherical.\cite{Beek:1967aa,Boyle:1985aa} The gradual increase in $n$, and accordingly $\varepsilon_{s}$, is consistent with a transition to more spherical mesoscale aggregates. The majority of the increase in $\varepsilon_{s}$ occurs over the 20-30mol$\%$ C$_{2}$MIm BF$_{4}$, with a plateau above $\approx$ 40mol$\%$ C$_2$MIm BF$_4$ implying that the non-polar domain continues to retain sphere-like morphology above this minimum concentration.

Due to the close agreement between the trends found by MD simulation, dynamic mechanical spectroscopy, and dielectric spectroscopy, we attribute the increase in $\varepsilon_{s}$, Figure \ref{fig: Figure4} (b), and accordingly the strength of the slow dielectric relaxation to a transition in the mesoscale aggregate morphology owing to the dilution of non-polar domains upon addition of C$_{2}$MIm BF$_{4}$. As a  direct result of the alteration in mesoscale aggregate morphology, the static dielectric permittivity of the ionic liquid mixtures are increased by almost 100$\%$ relative to the neat ionic liquids. 

The ability to tune $\varepsilon_{s}$ is significant since the vast majority of aprotic ionic liquids have low to moderate values of around 7-15, typical of low polarity solvents.\cite{Weingartner:2014aa,Huang:2011aa,Weingaertner:2008aa,Wakai:2005aa} Higher $\varepsilon_{s}$ values are expected to influence ionic liquid/solute and ionic liquid/solid-surface interactions with potentially critical implications for the application of ionic liquids in biomass processing, chemical synthesis, nanoparticle growth and as electrolytes in solar cells, batteries, and super-capacitors.\cite{Armand2009,Hallett:2011aa,Kim2012,Fedorov:2014aa} The substantial increase in $\varepsilon_{s}$ of the ionic liquid mixtures and its direct link to aggregate morphology and dynamics provides a new route to tuning this important physical parameter. More generally, this study highlights the advantage of combining techniques capable of probing composition-dependent changes in mesoscale aggregate morphology as well as mesoscale aggregate dynamics.

%The influence of mesoscale aggregate morphology and dynamics on the physicochemical properties of ionic liquid mixtures based on 1-octyl-3-methylimidazolium tetrafluoroborate (C$_{8}$MIm BF$_{4}$) and 1-ethyl-3-methylimidazolium tetrafluoroborate (C$_{2}$MIm BF$_{4}$) has been investigated by x-ray scattering, molecular dynamics simulations, dynamic-mechanical spectroscopy, and broadband dielectric spectroscopy. Upon dilution of the aggregating C$_{8}$MIm BF$_{4}$ with the non-aggregating C$_{2}$MIm BF$_{4}$ the bicontinuous non-polar domain is found to transition to increasingly isolated spherical non-polar aggregates. As a direct result of this transition, the dielectric strength of a slow dielectric relaxation attributed to interfacial polarization at the polar/non-polar domains is substantially increased resulting in a 100$\%$ increase the static dielectric permittivity, $\varepsilon_{s}$, of the mixtures relative to the neat parent ionic liquids. This demonstrates a new ability to tune the physicochemical properties of ionic liquids by the composition-dependent control of mesoscale aggregate morphology and dynamics afforded by simply by mixing two parent ionic liquids. It is envisioned that future mixture studies on a wide variety of chemical structures will yield ILs with additional and more complex self-assembled morphologies, the dynamics of which may produce ionic liquids with unprecedented and advantageous physical and chemical properties.

%%TC:ignore
J.S. acknowledges support by the National Science
Foundation, the Division of Chemistry through No. CHE-1753282. T.C. gratefully acknowledges financial support from the U.S. Army Research Office under Contract No. W911NF-17-1-0052. The BDS, DMS, and DSC measurements were conducted at the Oak Ridge National Laboratory's Center for Nanophase Materials Sciences, which is sponsored by the Division of Scientific User Facilities, Office of Basic Energy Sciences, U.S. Department of Energy. U. K. and J. K. S. acknowledge the support by the National Science Foundation. the Divison of Chemical, Biomolecular, Environmental, and Transport through the grant CBET-1706978.
The x-ray scattering experiments were performed at the Duke University Shared Materials Instrumentation Facility (SMIF), a member of the North Carolina Research Triangle Nanotechnology Network (RTNN), which is supported by the National Science Foundation (Grant No. ECCS-1542015) as part of the National Nanotechnology Coordinated Infrastructure
(NNCI). Sincere thanks to Prof. Austen Angell for helpful discussions. Computational resources were provided by the High Performance Computing Cluster at Oklahoma State University.
%%TC:endignore
\appendix

%\nocite{*}

\bibliography{bibliography}% Produces the bibliography via BibTeX.

\end{document}